\documentclass{tADP2e}

\usepackage{epstopdf}
\usepackage{subfigure}

\usepackage{upgreek}     

\newcommand{\micron}{~\ensuremath{\upmu\text{m}}}

\begin{document}


\articletype{Review}

\title{Laser spectroscopy of cold molecules}

\author{
\name{S. Borri and G. Santambrogio\thanks{$^\ast$Corresponding author. Email: santambrogio@lens.unifi.it. Present address Istituto Nazionale di Ricerca Metrologica, INRIM, Torino, Italy}}
\affil{Istituto Nazionale di Ottica-CNR \&
European Laboratory for Non-Linear Spectroscopy, LENS,
Via Nello Carrara 1,
50019 Sesto Fiorentino, Italy}
}

\maketitle

\begin{abstract}

This paper reviews the recent results in high-resolution spectroscopy
on cold molecules. Laser spectroscopy of cold molecules addresses
issues of symmetry violation, like in the search for the electric
dipole moment of the electron and the studies on energy differences in
enantiomers of chiral species; tries to improve the precision to
which fundamental physical constants are known and tests for their possible
variation in time and space; tests quantum electrodynamics, and
searches for a fifth force. 
Further, we briefly review the recent technological progresses in the
fields of cold molecules and mid-infrared lasers, which are the tools
that mainly set the limits for the resolution that is currently attainable
in the measurements.
\end{abstract}

\begin{classcode}06.20.-f Metrology, 06.30.Ft Time and frequency measurements, 33.20.-t Molecular spectra, 42.55.-f Lasers\end{classcode}

\begin{keywords}
Molecular spectroscopy, high-resolution spectroscopy, cold and
ultracold molecules, infrared light sources
\end{keywords}

\section{Introduction}
With the exception of a few dimers formed by the photo-association or
magneto-association of pre-cooled alkali-metal
atoms~\cite{Ni_PhysChemChemPhys11p9626_2009,Quemener_ChemRev112p4949_2012}, the ultracold world
(microkelvin and below) is
presently confined to atomic and ionic systems.
Recent spectroscopic measurements in atoms and atomic ions have
reached fractional accuracies of parts in
10$^{18}$.\cite{Nicholson_NatureComm6p6896_2015}
Precision measurements on ultracold atoms and atomic ions are at the
heart of the very best clocks in the world~\cite{Ludlow_RevModPhys87p637_2015},
magnetometers~\cite{Kominis_Nature422p596_2003},
gyroscopes~\cite{Berg_PhysRevLett114p063002_2015}, and
gravimeters~\cite{Fixler_Science315p74_2007}, and these devices are even
being developed into commercial products. 

The precision of spectroscopic studies on the molecular counterparts,
though, is worse by more than three orders of magnitude. This is a
consequence of the richer internal structure of molecules that makes
cooling and detection more complicated than in atoms. However, the
internal structure and symmetry of molecules, and the strong
intramolecular fields open the door to new kind of
measurements~\cite{Darquie_Chirality22p870_2010}. Spectroscopic
studies on cold molecules yield insight into new physics, in some
cases to a deeper level than using 
atoms despite the lower precision of the bare
measurements~\cite{Hudson_Nature473p493_2011,_Science343p6168_2013,Shelkovnikov_PhysRevLett100p150801_2008,Bagdonaite_PhysRevLett114p071301_2015,Bagdonaite_Science339p46_2013,Levshakov_AstrophysJ738p26_2011,Levshakov_AstronAstrophys512pA44_2010,Levshakov_AstronAstrophys524pA32_2010}. In 
some cases, laboratory measurements on cold molecules can even compete
in the search for new particles with the largest particle accelerator
facilities available.\cite{_Science343p6168_2013} The central part of
this paper will review the recent results in precision spectroscopy of
cold molecules.

The limits to the precision that is attainable in spectroscopic studies on
molecules are currently given by the cooling techniques and by laser
technology. Therefore, we will also briefly review here the recent
progresses in these two fields, from the perspective of
high-resolution molecular spectroscopy. 

\section{Techniques for the Production of Cold Molecules\label{coldmol}}

The shot-noise limit in a frequency measurement is given by $\delta 
\nu = 1/\tau \sqrt{N}$, where $\tau$ is the coherence time for the
measurement and $N$ is the number of detected molecules. The advantage
offered by using a cold sample of molecule is two-fold. First, by cooling
the internal degrees of freedom of a molecular sample, the
distribution of populated quantum states is narrowed according to the
Boltzmann statistics. Thus, $N$ is greatly enhanced if the state
under investigation is the ground state, or a state that is energetically
close to the ground state, compared with $k_\text{B}T$. As an aside,
we note that a current limit in molecular 
spectroscopy is posed by the lack of generally-applicable methods to
prepare a large population of molecules in highly excited
states. Second, cooling the external degrees of freedom yields slower
molecules that allow for measurements with longer coherence time
$\tau$. Moreover, lower velocities yield reduced Doppler broadenings.~\cite{Carr_NewJPhys11p055049_2009}

The coldest molecular species attainable to date are the ultracold
alkali dimers created by associating ultracold atoms.\cite{Quemener_ChemRev112p4949_2012} They have been
employed, for instance, in the studies of ultracold bimolecular
reactions~\cite{deMiranda_NaturePhys7p502_2011,Ospelkaus_Science327p853_2012}
and have been trapped in optical lattices to analyze their quantum
dynamics, which represents the first step towards
using these systems to explore many-body dynamics in regimes that are
inaccessible to current theoretical
techniques~\cite{Hazzard_PhysRevLett113p195302_2014}. However, as
interesting as these species are in the quest for new physical
phenomena, they have not been at the focus of the spectroscopists' attention.
Instead, their attention has been directed at less exotic molecular
species, cooled by direct methods. Here, we will briefly review the
main achievements in direct molecular cooling.

A conceptually simple method for preparing cold molecules is the
buffer-gas cooling pioneered by
J.~Doyle~\cite{Weinstein_Nature395p148_1998}. The cooling is achieved
via collisions with cryogenically cooled helium atoms and the
temperature of the cold molecules is typically around 1~K. A large
variety of atoms and molecules has been cooled using this technique,
ranging from atoms and dimers to benzonitrile, fluorobenzene,
anisole, for instance.\cite{Patterson_MolPhys110p1757_2012} 
Further, buffer-gas cooling has been used for the production of
molecular beams both of gaseous precursors and of laser ablated
species.\cite{Maxwell_PhysRevLett95p173201_2005} If curved guides are
coupled to this kind of sources, the 
subset of molecules that are moving sufficiently slowly are extracted
from the output of the
source.\cite{vanBuuren_PhysRevLett102p033001_2009}
Spectroscopy on buffer-gas cooled species has been performed both
inside the cooling
cell~\cite{Patterson_MolPhys110p1757_2012,Patterson_Nature497p475_2013,Spaun_Nature533p517_2016}
and on the molecular beam~\cite{_Science343p6168_2013}. 

Supersonic molecular beams are a classical method to produce cold molecular
samples. The first molecular beam was reported in
1911~\cite{Dunoyer_LeRadium8p142_1911} but is was only after Stern and
Gerlach's famous experiment in 1922~\cite{Gerlach_ZPhysik9p349_1922}
that this technique emerged as a planned scientific
effort.~\cite{Scoles1988}
A molecular beam is generated by letting a gas expand from a
high-pressure source into a low-pressure ambient background via some
sort of nozzle. In the expansion region of the beam, the initial
enthalpy of the gas is converted into forward velocity of the beam and
the temperature is consequently reduced. By seeding the molecules in
an inert gas, different final velocities can be obtained depending on
the average molecular weight of the gas mixture.
A supersonic molecular beam, thus, is a fast moving beam of internally
cold molecules and it facilitates spectroscopic studies by providing a
lack of collisional perturbations and the strong reduction of inhomogeneous 
broadening. 
However, typical molecular beams have speeds of the order of
300--1000~m/s and several techniques have been developed to
manipulate and control their motional degrees of freedom. 

Stark and Zeeman effects are used to manipulate polar and paramagnetic
molecules with electric and magnetic fields,
respectively.\cite{vanderMeerakker_ChemRev112p4828_2012,Lemeshko_MolPhys111p1648_2013}
Stark deceleration was first demonstrated by Gerard Meijer using
metastable CO~\cite{Bethlem_PhysRevLett83p1558_1999}. Molecules with
a permanent electric dipole moment convert part of their kinetic energy into
Stark energy upon entering an electric field if they are in an
appropriate quantum state. If the electric field is switched off
before the molecule has left the electric field, the lost kinetic
energy will not be returned. This process can be repeated over
multiple stages until the molecules reach the desired final
velocity. Once the average velocity is low enough, molecules can be
loaded in a trap~\cite{Bethlem_Nature406p491_2000}, for instance.
Zeeman deceleration is entirely analogous, except that the force is
exerted by a magnetic field on a magnetic dipole
moment.\cite{Vanhaecke_PhysRevA75p031402(R)_2007} 
Alternatively to the method of abruptly switching between 
different static field configurations, molecule can be captured in
traveling potential
wells~\cite{Meek_PhysRevLett100p153003_2008,Osterwalder_PhysRevA81p051401(R)_2010}
directly from the supersonic molecular beam and then decelerated.
These methods are used to prepare a molecular beam in a single quantum
state and at a mean speed adjustable between 400--500~m/s to rest,
with translational temperature tunable from 1~K to 5~mK. 
High-resolution spectroscopy of decelerated species has been performed
for NH$_3$~\cite{vanVeldhoven_EurPhysJD31p337_2004} and hydroxyl
radicals
(OH)~\cite{Hudson_PhysRevLett96p143004_2006,Lev_PhysRevA74p061402(R)_2006}. 
In these experiments, an interaction time as long as one millisecond was
obtained. 

In 2008, Stuhl et al.~\cite{Stuhl_PhysRevLett101p243002_2008}
identified a class of diatomic molecules that presented almost-cycling
transitions, which can be used for laser cooling. A couple of years
later the DeMille group demonstrated the action of a radiative force
acting on SrF~\cite{Shuman_PhysRevLett103p223001_2009}, and then
transverse laser
cooling~\cite{Shuman_Nature467p820_2010} and
deceleration~\cite{Barry_PhysRevLett108p103002_2012} of a SrF 
beam. Finally, in 2014, they were able to trap SrF in a
three-dimensional magneto-optical
trap~\cite{Barry_Nature512p286_2014}. In the meanwhile, laser cooling
of YO~\cite{Hummon_PhysRevLett110p143001_2013} and
CaF~\cite{Zhelyazkova_PhysRevA89p053416_2014,Hemmerling_arXivp1603.02787v1_2016}
has also been reported.  

Optoelectric cooling of polyatomics was first demonstrated by
Zeppenfeld et al.~\cite{Zeppenfeld_Nature491p570_2012} on
CH$_{3}$F. In a complementary fashion with respect to laser cooling, it
consists in a sequence of relatively few steps in each of which a large
fraction of a molecule's kinetic energy is removed. The energy is
extracted by allowing molecules to move in an electric field gradient
in different states with differing Stark energies. Spontaneous decay
provides the dissipation required to remove entropy. Very recently,
submillikelvin temperatures were obtained for 
an ensemble of 10$^{5}$ formaldehyde molecules.~\cite{Prehn_PhysRevLett116p063005_2016}

Typical temperatures reached with all these techniques are in the
range of a half~\cite{Norrgard_PhysRevLett116p063004_2016} to about a
hundred mK, whereas the densities are in the range of 10$^7$ molecules
per cubic centimeter. 

\section{Spectroscopic Studies with Cold Molecules}

\subsection{The Electric Dipole Moment of Fundamental Particles}

If a particle has an intrinsic electric dipole moment (EDM), the EDM
must necessarily lie along its spin axis because all other
perpendicular components would average out to zero. Time inversion (T)
would only reverse the direction of the spin, whereas parity inversion
(P) would only reverse the sign of the EDM. Therefore an EDM different
from zero leads to violations of T invariance (and
P)~\cite{Schiff_PhysRev132p2194_1963}. The CPT theorem states that the
combined operations of P, T, and charge conjugation (C) must be
conserved in any Lorentz-invariant
theory~\cite{Schwinger_PhysRev82p914_1951}. Moreover, in nearly all
current theories, violation of T implies a violation of CP symmetry. Indeed,
CP violation was first observed about 50 years ago in the decay of the
neutral kaon~\cite{Christenson_PhysRevLett13p138_1964} and such
violation can be explained through the standard model (SM). However,
the SM does not contain enough CP violation to explain the current
matter--antimatter asymmetry in the
universe~\cite{Balazs_PhysRevD71p075002_2005}, while also leading to
EDMs too small to be seen in any current or contemplated experiments.
Some theories that go beyond the SM generally provide for more CP
violation, and therefore, larger
EDMs.\cite{Pospelov_AnnPhys318p119_2005,Bernreuther_RevModPhys63p313_1991,Hoogeveen_NuclPhysB341p322_1990}
This makes the search for EDMs a powerful way to search for new
physics and constrains the possible extensions.  

Edward Purcell and Norman Ramsey initiated a search for an EDM of the
neutron over 60 years ago and obtained a result that is consistent with
zero.\cite{Purcell_PhysRev78p807_1950,Smith_PhysRev108p120_1957}
Thereafter, a long series of ever more sensitive EDM 
experiments began, on neutrons, atoms, and molecules. The neutron was
initially chosen because of the difficulties related with measurements
on charged particles. Moreover, in the
non-relativistic limit, an atom does not have an EDM even if the
electron does.\cite{Schiff_PhysRev132p2194_1963} But it was shown that
if relativity is taken into account, neutral atoms and molecules can
have an EDM~\cite{Sandars_PhysLett14p194_1965} and
this effect increases rapidly with the nuclear charge. Thus, 
experiments have been performed on atoms and molecules with heavy nuclei, like
Tl~\cite{Regan_PhysRevLett88p071805_2002} and
PbO~\cite{Eckel_PhysRevA87p052130_2013}. To date, the most accurate
EDM experiments measure the electron EDM in YbF and ThO
molecules~\cite{Hudson_Nature473p493_2011,_Science343p6168_2013}. This
larger accuracy is due to the polarizability of a typical polar 
diatomic molecule, which is about three orders of magnitude larger than
in an atom. This is reflected in the interaction energy of the
electron EDM with the electric fields inside these molecules, which
is hundreds of times larger than in
Tl.\cite{Kozlov_JPhysBAtMolOptPhys28p1933_1995} Moreover, the strong 
tensor polarizability of polar molecules greatly reduces the
systematic errors with respect to the measurements in
atoms.\cite{Hudson_PhysRevLett89p023003_2002} Efforts are currently
being made to cool further the molecular samples used in these
experiments in order to improve the sensitivity of the
measurements.\cite{Tarbutt_NewJPhys15p053034_2013} 
Furthermore, there has been notable progress by the Cornell group on
the measurements using trapped HfF$^+$ or ThF$^+$ molecular ions. By
applying a rotating bias electric field, they demonstrated effective
polarization of trapped molecular ions~\cite{Loh_Science342p1220_2013}.

Extensions of the SM that provide enough CP violation to explain the
matter--antimatter asymmetry do so by introducing new particles that
couple to the electron. The present experimental limit of sensitivity
is obtained with a cold sample of ThO molecules from a cryogenic buffer
gas beam source~\cite{_Science343p6168_2013} and
sets an upper limit of the electron EDM of $8.7\times
10^{-29}\;e\cdot$cm, which constrains the new supersymmetric
particles to masses larger than
TeV/$c^2$.\cite{Pospelov_AnnPhys318p119_2005,Bernreuther_RevModPhys63p313_1991} 
It is thus remarkable how laser spectroscopy on cold molecules can
compete in this field with the largest particle accelerator facilities
in the world.

\subsection{Parity Violation}

Another unresolved puzzle is about the overwhelming dissymmetry or
chirality of Earth biochemistry, which is based on L-amino acids and
D-sugars.\cite{Mason_Nature311p19_1984} While the efficiency of
homochiral chemistry and its stability toward natural 
selection is clear, little is known about the origin of this
particular choice. It might well be a pure matter of chance that the
initial enantiomeric distribution did not show an excess of D-ammino
acids and L-sugars, which might in fact be the chirality of the
carbon-based life on another planet. 

The origin of the observed chirality in biochemistry may be an effect
of the weak nuclear interaction. The weak interaction is unique in its
non-conservation of
P~\cite{Lee_PhysRev104p254_1956,Wu_PhysRev105p1413_1957} and this fact 
allows the effects of the weak interaction to be distinguished from
the much stronger, but P-conserving, electromagnetic interaction. In
chiral molecules, the energy shift caused by the weak
interaction changes sign from one molecule to its mirror
image.\cite{Hegstrom_JChemPhys73p2329_1980} Therefore, P does not
generate a true enantiomer because of the slight energy difference,
whereas the combined CP operation generates a mirrored molecule composed of
antiparticles and exactly the same energy. This follows from the CPT
theorem and the assumption that T is not
violated.\cite{Barron_MolPhys43p1395_1981,Barron_ChemPhysLett79p392_1981} 
It was suggested over 40 years ago that energy differences
between enantiomers should be measurable as differences in the
electronic~\cite{Rein_JMolEvol4p15_1974} and
vibrational~\cite{Letokhov_PhysLettA53p275_1975} energies of the two
enantiomers of a chiral molecule.
Further, cumulative amplification mechanisms have been proposed that
allow the tiny energy differences between enantiomeric molecules to
yield observable consequences for the chirality of our
biochemistry.\cite{Kondepudi_Nature314p438_1985} With such mechanisms, 
a minute but systematic chiral interaction can determine which
enantiomer will dominate in the long term. 

The first experiments probing P violation in chiral molecules
were performed on
CHFClBr~\cite{Daussy_PhysRevLett83p1554_1999,Ziskind_EurPhysJD20p219_2002},
using saturated-absorption laser spectroscopy in two Fabry-Perot
cavities containing samples with different enantiomeric excesses,
around a wavelength of 10\micron. In these experiments, the centers
of the absorption lines were determined to the Hz level, yielding an
upper bound of $\Delta\nu/\nu \sim 5 \times 10^{-14}$ for the P violation effect. 
However, theoretical
studies~\cite{Quack_JChemPhys119p11228_2003,Schwerdtfeger_PhysRevA71p012103_2005}
predict the line shift to be in the mHz 
range, corresponding to a precision of the order of parts in
$10^{16}$.
Therefore, new experiments are
planned~\cite{Darquie_Chirality22p870_2010} to improve the precision of
the measurements. The new generation of experiments is based on the
measurement of Doppler-free two-photon Ramsey fringes around
10\micron\ on a molecular 
beam~\cite{Shelkovnikov_PhysRevLett100p150801_2008}. 

In diatomic free radicals, it is possible to bring two states of opposite parity to near degeneracy by inducing a Zeeman shift as large as the rotational splitting. Near such a level crossing, the mixing of these long-lived states due to nuclear spin-dependent P-violating interactions is greatly enhanced~\cite{Flambaum_PhysLettA110p121_1985}. It has been suggested that these systems could be used to measure classes of P-violating electroweak interactions that are difficult to access otherwise\cite{Flambaum_PhysLettA110p121_1985,DeMille_PhysRevLett100p023003_2008}, such as those due to nuclear anapole moments and axial hadronic-vector electronic electroweak couplings~\cite{Flambaum_PhysRevC56p1641_1997}. Results in this direction are expected by ongoing experiments in the DeMille and in the Hoekstra groups~\cite{Cahn_PhysRevLett112p163002_2014,vandenBerg_JMolSpectr300p22_2014}. 

\subsection{Variations of fundamental constants}

The search for EDMs of fundamental particles and the measurement of P
violation in chiral molecules question the extent to which the
symmetry-conservation rules postulated in the SM hold. Another aspect of the SM that is
presently under scrutiny is whether its fundamental constants are
fixed parameters, or are rather changing over time, in space, or in
dependence of matter density. This issue is not of secondary
importance, as the comparison and reproduction of experiments is at
the foundation of the scientific method, but it is only meaningful
if the natural laws do not depend on time and space (Einstein's
equivalence principle). 

The question of a possible variation of the fundamental constants was
probably first posed by Dirac in the
30s~\cite{Dirac_Nature139p323_1937}, and since then many other theories
have been
developed.\cite{Uzan_RevModPhys75p403_2003,Uzan_LivingRevRelativ12p2_2011}
Moreover, the observations of the last 20 years that indicate that the
universe is expanding at an accelerating
rate~\cite{Perlmutter_RevModPhys84p1127_2012,Schmidt_RevModPhys84p1151_2012,Riess_RevModPhys84p1165_2012}
leads to the postulation of an unknown form of energy, known as dark
energy. Two proposed hypotheses for dark energy are the cosmological
constant and the dynamical action of a scalar
field.\cite{Ratra_PhysRevD37p3406_1988} For the latter case, it has
been shown that the scalar field must interact with matter, giving
rise to a variation of the fundamental coupling
constants~\cite{Carroll_PhysRevLett81p3067_1998,Bekenstein_PhysRevD66p123514_2002} and reinvigorating the interest for this field of research.

While models of the big bang nuclear synthesis set limits on the
variation of fundamental constants at extremely high
redshifts~\cite{Berengut_PhysRevD87p085018_2013}, the measurement of
atomic and molecular transition frequencies is the most
natural way to look for variation of the fine structure constant,
$\alpha$, and the electron-to-proton mass ratio, $\mu$, from intermediate
redshifts ($z\sim 5$) to the current epoch. Observation of
intergalactic species found in the line-of-sight of quasars yields
sufficient spectral quality up to about
$z=4$~\cite{Bagdonaite_PhysRevLett114p071301_2015}, whereas
measurements of samples from within the Milky Way can test the
hypothesis that fundamental constants may differ between the high- and
low-density environments of the Earth and the interstellar
medium~\cite{Truppe_NatureComm4p2600_2013}, or between the
gravitational potential of white dwarfs and the
Earth~\cite{Salumbides_MNRAS450p1237_2015}. 

Of course, it is essential that the
different transitions being compared have different dependency on
$\alpha$ and $\mu$. Amy-Klein and co-workers, for example, compared a
vibrational transition in SF$_6$, which depends directly on $\mu$, to
a hyperfine transition in a Cs clock, which depends on $\alpha$
instead.\cite{Shelkovnikov_PhysRevLett100p150801_2008} The comparison
of different clock transitions in ultracold atoms or atomic ions
provides a high signal-to-noise ratio and can be carried out under
very well controlled
conditions~\cite{Rosenband_Science319p1808_2008}. Alternatively, one
can choose systems that are not necessarily ideal for precision
measurements, but present an enhanced sensitivity (up to three orders
of magnitude) to a variation of physical constants. Several molecular
systems have been proposed in which a near degeneracy between
electronic levels of different
symmetry~\cite{DeMille_PhysRevLett100p043202_2008}, between hyperfine
and rotational levels~\cite{Flambaum_PhysRevA73p034101_2006}, or
between fine structure and vibrational
levels~\cite{Flambaum_PhysRevLett99p150801_2007,Bethlem_FaradayDiscuss142p25_2009},
leads to particularly large sensitivities.

Currently, all relevant experimental results on the variation of $\mu$ are obtained
by measuring molecular transitions. The present-day limit is set in
the laboratory with SF$_6$ at $\dot{\mu}/\mu=(-3.8\pm5.6)\times
10^{-14}\,\text{yr}^{-1}$~\cite{Shelkovnikov_PhysRevLett100p150801_2008};
the comparison of modern measurements of the Lyman and Werner band of H$_2$
and the astrophysical observations from 12.4 billion years ago yields
$\Delta \mu/\mu=(-9.5\pm5.4_\text{stat}\pm5.3_\text{syst})\times
10^{-6}$~\cite{Bagdonaite_PhysRevLett114p071301_2015}; a similar study
for methanol using a radio-telescope to look back 7 billion years
yields $\Delta \mu/\mu=(-0.0\pm1.0)\times
10^{-7}$~\cite{Bagdonaite_Science339p46_2013}. Measurements within our
galaxy, testing whether constants depend on the local density, are
available for methanol maser lines and set an upper limit of $|\Delta
\mu/\mu|=2.8 \times 10^{-8}$
(1$\sigma$).\cite{Levshakov_AstrophysJ738p26_2011} However,
comparisons of terrestrial and astrophysical microwave transitions in
ammonia and other molecules, find an eight-standard-deviation
systematic difference. This suggests a fractional change in $\mu$ of
$2.6\times 10^{-8}$ when going from the Earth to the interstellar
medium, tentatively supporting the chameleon
hypothesis.\cite{Levshakov_AstronAstrophys512pA44_2010,Levshakov_AstronAstrophys524pA32_2010} 
Also the strongest limit on the dependency of $\alpha$ on matter density has been obtained with a molecular measurement, CH~\cite{Truppe_NatureComm4p2600_2013}. This was done by measuring microwave transitions in CH and by comparing these frequencies with those measured from sources of CH in the Milky Way. 
High-precision measurements of the most sensitive molecular transition
frequencies are thus required, together with higher-quality
astronomical observations. 




\subsection{Test of QED and Fifth Force}
A further approach to probe experimentally new physics is the search
for a new attractive or repulsive force (a \emph{fifth force}) with
precision metrology measurements on calculable molecular systems. In
recent years, thorough level structure calculations, including QED and
high-order relativistic contributions, have been carried out for the
neutral hydrogen molecules and the deuterated
isotopologues~\cite{Piszczatowski_JChemTheoryComput5p3039_2009,Pachucki_PhysChemChemPhys12p9188_2010,Pachucki_JChemPhys130p164113_2009,Komasa_JChemTheoryComput7p3105_2011}. In
these systems, the energy level shifts due to weak
interaction and gravity are orders of magnitude away from the experimental
sensitivity. The effect of the strong interaction is confined to the
fm scale and its influence on atomic and molecular energy levels
enters into the calculations via the nuclear $g_N$ factor and the
nuclear spin. Thus, a search for deviation from the QED predictions in
atomic and molecular energy levels would either hint at a new kind of
interaction or at some unaccounted effects within QED. Furthermore,
whereas lepton-nucleon and lepton-lepton interactions may be probed in
atomic hydrogen and helium, the search for long-range interactions
between hadrons requires a molecular
system.\cite{Salumbides_PhysRevD87p112008_2013} The measurements with
the highest precision to date are in perfect agreement with the
calculation~\cite{Salumbides_PhysRevLett107p043005_2011}. However, the
accuracy level of QED calculations is claimed to be one order of
magnitude better than the present
experiments~\cite{Komasa_JChemTheoryComput7p3105_2011}, thus improved
experimental tests are currently required.

\subsection{Determination of the Boltzmann Constant}
The Boltzmann constant, $k_\text{B}$, conventionally considered as
fundamental, plays an important role for a possible redefinition of
the kelvin, one of the seven SI base
units~\cite{Flowers_RepProgPhys64p1191_2001}. Currently, the most
precise determination of $k_\text{B}$ comes from measurements of the
speed of sound in a noble gas inside an acoustic
resonator~\cite{dePodesta_Metrologia50p354_2013}. The measured value
is $k_\text{B}$ = 1.380 651 56 (98)$\times$10$^{-23}$ J$\cdot$K$^{-1}$,
with a relative uncertainty of 0.71$\times$10$^{-6}$. Thanks to recent
developments of frequency-stabilized ultra-narrow coherent sources,
precise determination of the Boltzmann constant beyond the
10$^{-6}$-level via spectroscopic measurements becomes a concrete
target. This method, called Doppler broadening thermometry
(DBT), consists in retrieving the Doppler width of a given atomic or
molecular line in a gas sample at thermodynamic equilibrium by
highly accurate spectroscopic detection of the line profile. The first
DBT experiment was performed at LPL with an ultra-narrow CO$_2$
laser~\cite{Daussy_PhysRevLett98p250801_2007}. The thermometric gas
employed was ammonia, kept in thermal water-ice bath at 273.15~K. The
Doppler width of an ammonia absorption line around 10.36\micron,
extrapolated at zero gas pressure, allowed for $k_\text{B}$
determination with a relative uncertainty of
2$\times$10$^{-4}$. Critical factors affecting the measurement come
from both the spectroscopic apparatus (stability and homogeneity of
the gas temperature in the spectroscopic cell, frequency and amplitude
stability of the laser source, linearity and noise level of the
detection chain) and data analysis (line-shape modeling and fitting
procedure). Similar results have been obtained with DBT measurements
on a CO$_2$ transition around 2\micron\ using an extended-cavity diode
laser~\cite{Casa_PhysRevLett100p200801_2008}, while a precision of
2.4$\times$10$^{-5}$ has been achieved on H$_2\,^{18}$O transitions
at 1.39\micron~\cite{Moretti_PhysRevLett111p060803_2013} mainly thanks
to a more refined line shape model. Significant improvements can be
expected both by a better laser stabilization and by adopting a more
sophisticated line-shape model for highly accurate retrieval of the
Doppler width~\cite{Rohart_PhysRevA90p042506_2014}.

\section{Laser sources in the Mid IR}

The most straightforward spectral region for high-resolution
spectroscopy are the microwaves, where radiation sources are very
reliable, stable, easy to use, and powerful; Doppler-broadenings
are tiny; and spatial coherence is easily achieved over the typical
sizes of an experimental apparatus. However, for a given molecular velocity
(i.e. translational temperature), the interaction time of a molecule with
the radiation is limited, thus increasing the frequency of the
measured transition allows for an improvement in the relative precision of the
measurement. The mid IR (MIR) corresponds to the frequencies of nuclear
vibrations and is, thus, a natural spectral region for molecular
studies, in which one finds intense rovibrational transitions,
accompanied by Hz-level natural linewidths. The larger Doppler effect
(which is proportional to the radiation frequency) must be dealt with
an intrinsic sub-Doppler spectroscopic technique, like two-photon
spectroscopy or saturated-absorption spectroscopy. Therefore, moving
from the GHz to the THz frequency range can produce a dramatic increase in 
the precision but it requires the development of intense and narrow-linewidth
laser sources to be referenced to the primary microwave
standard.
In other words, the precision of a molecular spectroscopic measurement is the result of the efforts towards the cooling of the molecules, which are reviewed in Section~\ref{coldmol}, and towards the improvement of the light sources, mainly in terms of frequency stability, which are reviewed below. These two technological fields are somehow complementary. Doppler broadening can be reduced either by slowing down the molecular motion or by adopting a sub-Doppler spectroscopic technique, if the available laser intensity is sufficient. Similarly, the coherence time is improved with slower molecules, and a more intense laser allows for a larger beam waist and thus for a longer interaction time.

Indeed, measurements of absolute frequencies must ultimately be
referenced to the primary frequency standard, which is based on the
hyperfine ground-state splitting in cesium. The comparison of the
measured transition frequencies in the optical domain (hundreds of
THz) to the microwave cesium frequency standard (around 9~GHz) was a
significant technical challenge in the past. However, the development
of optical frequency combs (OFC) has allowed to bridge this
four-orders-of magnitude gap directly, leading to measurements of unprecedented
precision~\cite{Hansch_RevModPhys78p1297_2006,Hall_RevModPhys78p1279_2006}. Nowadays
OFCs are commercially available in the visible-to-near-IR (VIS-to-NIR) region but
remain challenging in the mid IR. Similarly, narrow-linewidth laser
sources, which are the other fundamental ingredient for
high-resolution spectroscopy and a mature technology in the
VIS-to-NIR, are still in their infancy in the MIR. 

One possible approach to the absolute determination of MIR frequency
is based on a two-step strategy. An OFC transfers the primary frequency
standard to the NIR, then difference-frequency generation (DFG)
from two sources, which are both referenced directly to the comb,
provides light in the MIR. For
this to be possible, the DFG pump and signal lasers must fall in the
OFC coverage range and their frequencies must be locked to the
nearest teeth of the comb. This leads to a very narrow idler
linewidth, only limited by the excess phase noise between the two comb
teeth due to the propagation of the repetition rate phase-noise to the
optical frequencies. The first comb-assisted DFG sources have been
used to measure the frequency of some CO$_2$ transitions around
4.3\micron\ by cavity-enhanced saturated-absorption 
spectroscopy, achieving an uncertainty of 800 Hz in the absolute
frequencies (1.1$\times$10$^{-11}$ relative precision).\cite{Mazzotti_OptLett30p997_2005}
Similar approaches have been adopted by other groups~\cite{Bressel_OptLett37p918_2012} covering
different MIR spectral ranges. Further, if the free-running short-term
stability of at least one of the DFG pumping lasers is better than the
comb's, this scheme can be improved: the $n/m$ excess phase noise
between the $n^\text{th}$ and $m^\text{th}$ tooth to which the pump and signal lasers
are locked can be canceled out using a direct digital
synthesizer.\cite{Telle_ApplPhysB74p1_2002} 
With this approach, absolutely-linked idler radiation between 4 and
4.5\micron, with 10~Hz
intrinsic linewidths (1~kHz integrated linewidth over 1 ms), has been
demonstrated~\cite{Galli_OptExpress17p9582_2009}. Entirely analogous
is the generation of OFC-referenced MIR radiation using optical
parametric oscillators
(OPOs)~\cite{Kovalchuk_OptLett30p3141_2005,Vainio_OptLett36p4122_2011,Ricciardi_OptExpress20p9178_2012}. 

An alternative approach consists in down-converting
a visible or NIR OFC directly to the MIR\cite{Schliesser_NaturePhoton6p440_2012}, either by
using a DFG processes
\cite{Zimmermann_OptLett29p310_2004,Maddaloni_NewJPhys8p262_2006,Enry_OptLett32p1138_2007,Gambetta_OptLett33p2671_2008,Phillips_OptLett37p2928_2012,Gambetta_OptLett38p1155_2013}
or an OPO\cite{Adler_OptLett34p1330_2009,Leindecker_OptExpress19p6296_2011,Coluccelli_OptExpress20p22042_2012,Ulvila_OptExpress22p10535_2014}. 
DFG MIR combs benefit from being offset free, because of the
perfect cancellation of any carrier-envelope phase offset that may be
present in the original frequency
comb.\cite{Galli_OptExpress21p28877_2013} The typical total
average power of a DFG OFC is in the order of a few mW, which results
in extremely low power per tooth. Therefore the application of DFG
OFCs as sources for direct high-resolution spectroscopy is not
common. In OPO MIR combs, the optical cavity is highly reflective for idler
wavelengths, and matched in length to the pump (usually a mode-locked
fiber laser) repetition rate. Methods for controlling the offset
frequency and mode spacing of the frequency comb have also been
demonstrated~\cite{Ulvila_OptExpress22p10535_2014}. These combs yield an average W-level
power, but their main drawbacks are the additional complexity given by
the OPO cavity, the relatively limited oscillation spectral range, and
the need for complex techniques for phase stabilization of both signal
and idler outputs.

It is noteworthy that the limited availability of suitable
nonlinear crystals often seriously restricts the possible choices for
pump/signal/idler combination, according to the transparency range of
the material or to the phase-matching requirements. For these reasons,
many of the cited DFG and OPO combs fall in the 2.5--4.5\micron\ range,
where PPLN crystals are transparent. Moreover, a proper choice of the
poling period allows for quasi-phase-matching with Nd:YAG
lasers at 1.064\micron, which is often convenient due to their high
stability and power levels. It is more difficult to access the region
above 5\micron, where PPLN crystals are not transparent. Here, other
crystals are commercially available (AgGaSe, 
AgGaS$_2$, GaSe, ZnSeP$_2$), but present low conversion
efficiencies and their transparency ranges and phase-matching requirements
strongly limit the choice of pump, signal and idler
sources. Some non-commercial crystal can be used, such as CdSeP$_2$ or
the orientation-patterned (OP)-GaAs or -GaP. These crystals are
characterized by high conversion efficiencies (at the level of PPLN or
higher). Bulk crystals like CdSeP$_2$ allow for a wide spectral
coverage, while OP-crystals are more selective according to the
patterning period. 

A third approach consists in the use of a relatively
high-power MIR laser, emitting at the desired frequency, stabilized
over some narrow spectral feature (high-finesse optical resonator, sub-Doppler
molecular transition, narrow-linewidth optical reference, for
instance) that is, in turn, referenced to a frequency standard.
Researchers at LPL in Paris, for example, stabilized CO$_2$ lasers on
a Fabry-Perot cavity filled with OsO$_4$ to achieve a 10-Hz-level
linewidth, a 0.1~Hz stability over 100~s and a reproducibility up to
10~Hz.\cite{Bernard_IEEEJQuantElec33p1282_1997} Then, to determine the
absolute frequency of their laser, they produced sum frequency (SF)
radiation in a AgGaS$_2$ crystal of the CO$_2$ laser and a visible
laser that is referred to an OFC. The SF radiation is also visible and
can be measured against the same OFC.\cite{AmyKlein_OptLett30p3320_2005} 

The lack of tunability and limited spectral coverage of gas lasers, however,
severely limits the range of molecules that can be studied. It is
mainly for this reason that the introduction of Quantum Cascade Lasers
(QCLs) has revolutionized the field of MIR
spectroscopy. They allow for continuous tunability over tens of
wavenumbers and a complete coverage, by 
design, of the MIR spectral range from 4 to 
20\micron, at
least~\cite{Maulini_ElectronicLetters45p107_2009}. Moreover, their output power has been
demonstrated to reach the W
level~\cite{Lu_ApplPhysLett98p181106_2011}.
In 2007, the frequency of a free running QCL around 4.3\micron\ was absolutely
referenced to a comb operating between 500 and 1100~nm by SFG of the
QCL radiation with the fundamental of a Nd:YAG in a PPLN
crystal.\cite{Bartalini_OptLett32p988_2007,Borri_OptExpress16p11637_2008} Analysis of the noise
features showed that QCLs have very narrow intrinsic
linewidths.\cite{Borri_IEEEJQuantElec47p984_2011,Bartalini_PhysRevLett104p083904_2010,Bartalini_OptExpress19p17996_2011}
This observation triggered a series of improvements in stabilization
of the QCL chip temperature and driving current. Thereafter, the
frequency of QCLs was locked to a sub-Doppler
transition~\cite{Cappelli_OptLett37p4811_2012}, to MIR Fabry-Perot
cavities~\cite{Taubman_OptLett27p2164_2002,Fasci_OptLett39p4946_2014},
to a OsO$_4$-stabilized CO$_2$
laser~\cite{Sow_ApplPhysLett104p264101_2014}, to a crystalline
whispering gallery mode
microresonator~\cite{SicilianideCumis_LaserPhotRev10p153_2016}, and to a
NIR ultra-low expansion reference cavity after up-conversion by
SFG~\cite{Hansen_OptLett40p2289_2015} as references. A linewidth as
narrow as 10~Hz with a relative stability in the 10$^{-14}$ range at
1~s, and a relative accuracy of 3$\times$10$^{-12}$, was
demonstrated~\cite{Sow_ApplPhysLett104p264101_2014}. 
Phase locking of QCLs to OFCs has been done with the various up- and
down-conversion strategies described
above~\cite{Mills_OptLett37p4083_2012,Hansen_OptExpress21p27043_2013,Gambetta_OptLett40p304_2015,Galli_OptExpress21p28877_2013},
and via optical injection locking~\cite{Borri_OptLett37p1011_2012},
thereby allowing also for the narrowing of the linewidth ranging from tens of
kHz down to a few hundreds of Hz. The most spectacular result to-date
was reported by LPL group (sub-Hz linewidth) by locking a QCL at
10.3\micron\ to an OFC, which is itself stabilised to a remote NIR
ultra-stable frequency reference via an optical fibre~\cite{Argence_NaturePhot9p456_2015}.

At these levels of precision, the issue of dissemination of the
primary frequency standard becomes of paramount importance. 
Standard OFCs are actively stabilized against a 10-MHz
quartz-oscillator disciplined by a Rb-GPS (Global
Positioning System) clock. The GPS stability and the quartz oscillator
phase noise limit the comb stability to parts in 10$^{13}$ in 1~s
and the absolute accuracy to the 10$^{-12}$ level.\cite{Niering_PhysRevLett84p5496_2000}
To overcome this limit, the primary frequency standard is now
delivered to some laboratories by the national metrological institutes
directly via fiber link, improving by more than four orders of
magnitude the resolution of satellite transfer techniques
\cite{Bauch_Metrologia43p109_2006}. Among these, are the PTB in Germany, the LNE-SYRTE
in France, the INRIM in Italy, the AGH in Poland, and the NPL in the
UK, transferring the accuracy and resolution of their 
atomic clocks for hundreds of kilometers.
Frequency instability of 3$\times$10$^{-19}$ over 1000~s were
measured, with ultimate accuracies on the frequency transfer of parts
in 10$^{19}$ (1000~s integration
time)~\cite{Calonico_ApplPhysB117p979_2014}. 
In fact, it was thanks to a fiber link from LNE-SYRTE that at LPL it
was possible to measure the absolute accuracy of a QCL at the
10$^{-14}$ level~\cite{Argence_NaturePhot9p456_2015}.

\section{Perspectives}

With precision measurements on atoms reaching a total uncertainty of
parts in 10$^{18}$~\cite{Nicholson_NatureComm6p6896_2015}, the gap
with the precision of molecular measurements is about three orders of
magnitude. The main reason for this poor performance is
arguably that molecular samples on which one wants to do
spectroscopy are much warmer, at the mK level at best, corresponding
to velocities of the order of the meter per second, depending on the
molecular mass. This constrains the interaction time with laser light
to the millisecond range, assuming a laser beam waists of the order of the
millimeter, which is optimistic in the case of a two-photon process
required for sub-Doppler spectroscopy. One can push this limit with a
Ramsey interrogation scheme, either in a
beam~\cite{Shelkovnikov_PhysRevLett100p150801_2008,Hudson_Nature473p493_2011,_Science343p6168_2013}
or building a fountain~\cite{Bethlem_EurPhysJSpecialTopics163p55_2008}.
Yet this comes at the cost of lower number densities,
larger setups, and, thus, worse control on stray fields.
Therefore, the development of a second-stage cooling method is
currently one of the biggest challenges in the field. One of the most
promising proposals is sympathetic cooling, which is based on the conceptually
simple idea of bringing cold molecules into thermal contact with a
bath containing ultracold atoms. So far sympathetic cooling has been
successfully accomplished for
ions~\cite{Larson_PhysRevLett57p70_1986,Ostendorf_PhysRevLett97p243005_2006}
and some neutral
atoms~\cite{Myatt_PhysRevLett78p586_1997,Modugno_Science294p1320_2001},
but not for neutral molecules. 

Another challenge in the field is the extension of cooling and highly
sensitive detection techniques to complex, polyatomic
molecules. Presently, cooling these systems to a few kelvin and forming
slow-moving beams would vastly extend the range of molecules that can
be brought under control to enable high-precision
measurements. 

Finally, MIR sources present a twofold challenge: on one side,
extending their spectral coverage to the entire MIR window from 3
to 25\micron\ allows to bring accurate frequency metrology methods to
almost all molecules; on the other side, improving the light source
stability and accuracy to the 10$^{-15}$ level, at least, is
required for many of the experiments described above. In this regard, 
the recent results with ultra-narrow, fiber-link-referenced MIR
sources have just provided a major improvement.

\section*{Acknowledgments}
We thank Davide Mazzotti for his helpful comments. 

\section*{Funding}
We gratefully acknowledge funding by INFN under the SUPREMO project,
by LASERLAB-EUROPE under grant agreement 284464, and by Extreme Light
Infrastructure (ELI) European project .

\bibliographystyle{gams-notit-nonumb}
\bibliography{bib}

\end{document}